\documentclass[twocolumn,prb,superscriptaddress,amsmath,amssymb,showpacs]{revtex4}

\usepackage{graphicx}

\renewcommand{\Re}{\mathop\mathrm{Re}\nolimits}

 \DeclareMathOperator{\TR}{\mathbf{Tr}}
 \DeclareMathOperator{\Tr}{Tr}
 \DeclareMathOperator{\tr}{tr}
 \DeclareMathOperator{\sgn}{sgn}

\begin{document}

\title{Proximity effect in the presence of Coulomb interaction and magnetic field}

 \author{P.~M.~Ostrovsky}
 \email{ostrov@itp.ac.ru}
 \affiliation{Institut f\"ur Nanotechnologie, Forschungszentrum Karlsruhe, 76021 Karlsruhe, Germany}
 \affiliation{L.~D.~Landau Institute for Theoretical Physics RAS, 119334 Moscow, Russia}

 \author{Ya.~V.~Fominov}
 \email{fominov@landau.ac.ru}
 \affiliation{L.~D.~Landau Institute for Theoretical Physics RAS, 119334 Moscow, Russia}

 \author{M.~V.~Feigel'man}
 \email{feigel@landau.ac.ru}
 \affiliation{L.~D.~Landau Institute for Theoretical Physics RAS, 119334 Moscow, Russia}

 \date{14 September 2006}

\begin{abstract}
We consider a small metallic grain coupled to a superconductor by a tunnel contact. We study the interplay between
proximity and charging effects in the presence of the external magnetic field. Employing the adiabatic approximation we
develop a self-consistent theory valid for an arbitrary ratio of proximity and Coulomb strength. The magnetic field
suppresses the proximity-induced minigap in an unusual way. We find the phase diagram of the grain in the charging
energy -- magnetic field plane. Two distinct states exist with different values and magnetic field dependences of the
minigap. The first-order phase transition occurs between these two minigapped states. The transition to the gapless
state may occur by the first- or second-order mechanism depending on the charging energy. We also calculate the
tunneling density of states in the grain. The energy dependence of this quantity demonstrates two different gaps
corresponding to the Coulomb and proximity effects. These gaps may be separated in sufficiently high magnetic field.
\end{abstract}



\maketitle

\section{Introduction}

A normal metal in contact with a superconducting lead acquires some superconducting properties; this phenomenon is known
as the proximity effect (for review see, e.g., Ref.~\onlinecite{S&M}). In particular, the electron spectrum of the
normal metal becomes gapped; at low interface transparency this gap is much smaller than superconducting $\Delta$, hence
the name minigap. This phenomenon is due to the Cooper pairs of the superconductor that penetrate into the normal metal
and induce weak superconductive correlations there.

If a normal metal part of the superconductor -- normal metal junction is a small grain, then the Coulomb effects come
into play (adding an electron to the grain costs charging energy). They are mostly pronounced in tunneling experiments
when the differential conductance between the normal external tip and the small grain is measured [this quantity is
proportional to what is called the tunneling density of states (TDOS)]. The Coulomb repulsion between tunneling
electrons reduces the current. This phenomenon is known as the tunneling anomaly or, in the zero-dimensional case,
Coulomb blockade.\cite{GD} Except for the charging energy, an essential parameter governing the efficiency of the
Coulomb blockade is the interface conductance $G$ between the grain and the lead. In the case of a normal lead the
Coulomb blockade is developed\cite{GD} at $G \ll 1$ (we measure $G$ in units of $e^2/\hbar$) and disappears\cite{WG} at
$G\gg 1$ due to the fact that an electron tunneling to the grain is rapidly transferred to the lead, thus not blocking
the tunneling of the next electron. At the same time, the Coulomb blockade persists even at $G\gg 1$ if the lead is
superconducting because a single electron cannot escape into the lead due to the gap $\Delta$ in its single-particle
density of states (DOS). This situation was studied by Matveev and Glazman.\cite{MG}

Apart from the tunnel current, the Coulomb interaction suppresses the proximity effect as well. The charging energy
prevents Cooper pairs from tunneling to and from the normal grain and thus destroys the superconducting order induced on
the normal side of the junction. Qualitatively, the Coulomb interaction is trying to fix the charge (neutrality) of the
normal grain while the proximity effect fixes the phase. Charge and phase are conjugate variables obeying the
uncertainty principle: they may not be fixed simultaneously. Recently, a full quantitative description of this
competition between the proximity and Coulomb effects in superconductor -- normal metal junctions at large interface
conductance was derived in Ref.~\onlinecite{OSF}. Obviously, the Coulomb interaction diminishes the minigap. In this
paper we extend the results of Ref.~\onlinecite{OSF}, including the external magnetic field into consideration. The
magnetic field can easily be varied in experiment, and we demonstrate that the minigap is qualitatively sensitive to the
strength of the magnetic field.

We consider a normal grain connected to a superconducting lead by tunnel junctions (low interface transparency) with
large conductance $G$ (determined by the product of transparency by the number of channels). We assume the
zero-dimensional limit; i.e., the Thouless energy $E_\mathrm{Th} = D/d^2$ is larger than all other relevant energy
scales of the system including the superconductive gap $\Delta$ in the leads (here $D$ is the diffusion constant in the
grain and $d$ is its characteristic size). In this limit, the proximity-induced minigap\cite{GK} is $E_g = G\delta/4$,
provided $E_g \ll \Delta$ and $\delta$ being the mean level spacing per one spin projection in the grain.

Our aim is to take into account effects of Coulomb interaction and magnetic field. The characteristic Coulomb energy
$E_C = e^2/2C$ is assumed (similarly to Ref.~\onlinecite{OSF}) to lie in the same range as $E_g$:
\begin{equation} \label{conditions}
\Delta \gg (E_g, E_C) \gg \delta.
\end{equation}
The capacitance $C$ of the grain already takes into account the renormalization $C = C_0 + e^2 G/2\Delta$ due to virtual
quasiparticle tunneling.\cite{LO} This renormalization assures the inequality $E_C \ll \Delta$ and allows arbitrary
small geometric capacitance $C_0$.

We consider relatively weak magnetic fields $H$ that do not affect the superconducting lead; i.e., $H$ should be much
smaller than the critical field of the lead: $H\ll H_{c2} = \Phi_0 / 2\pi \xi_S^2$, where $\Phi_0$ is the flux quantum
and $\xi_S = \sqrt{D_S/\Delta}$ is the superconductive coherence length. Note that since the paramagnetic limit is much
larger than $H_{c2}$, the condition $H \ll \Delta$ is certainly fulfilled (here and below we express $H$ in energy units
dropping the factor $g\mu_B /2$).

If the normal grain is sufficiently small, we can neglect the orbital effect of the magnetic field in the grain. Indeed,
as $d$ decreases, the critical field due to orbital effects in a superconductor with a gap $E_g$ grows as $\Phi_0/ \xi
d$, where $\xi = \sqrt{D/E_g}$ is the coherence length corresponding to the proximity-induced superconductivity. At the
same time, the Zeeman effect of the magnetic field determines the paramagnetic limit with the critical field of the
order of $E_g$ independent of the grain's size. Hence the disregard of the orbital effect is justified for small grains
with $d \ll \Phi_0/ \sqrt{D E_g}$.

To take into account both the proximity and charging effects, the adiabatic approximation was employed in
Ref.~\onlinecite{OSF}. An inequality $E_C \gg \delta$ provides the separation of energy scales: electronic degrees of
freedom (which are contained in the matrix $Q$ of the $\sigma$ model; see below) are ``slow'' compared to the
characteristic frequency of the electric potential fluctuations. This allows one to calculate the renormalized (due to
interaction) value of the minigap $\tilde E_g$. The result of the competition between charging and proximity effects is
determined by a comparison of the charging energy $E_C$ with the effective ``Josephson'' energy $E_J \propto G^2 \delta
\ln(\Delta/E_g)$. The latter has a clear meaning of the Josephson coupling energy\cite{AB} between the superconductive
reservoir and an imaginary weak superconductor with the order parameter $E_g$. The two limiting cases of the Coulomb
versus proximity competition are the (i)~\emph{weak Coulomb blockade} limit $E_J \gg E_C$, when a small negative
correction to the noninteracting minigap $E_g$ arises, and (ii)~\emph{strong Coulomb blockade} regime $E_J \ll E_C$,
when the minigap is exponentially suppressed. A self-consistent approach allowing for the magnetic field is developed in
Sec.~\ref{sec:model}.

As we show below, the magnetic field does not influence the minigap $\tilde E_g$ if $H < \tilde E_g /2$, while in the
opposite case extra solutions of the model appear, resulting in a rich phase diagram describing different possible
values of the minigap. An extensive study of these solutions is presented in Sec.~\ref{sec:minigap}.

The TDOS is also strongly affected by the magnetic field. This quantity is particularly interesting because it can be
directly measured in the experiment as the differential conductance between the normal grain and a normal external tip
--- e.g., with the help of the tunneling microscopy technique. The Coulomb interaction produces a drastic impact on the
TDOS: as the charging energy increases, the proximity minigap gradually transforms into a Coulomb gap of the order of
$E_C$. The magnetic field significantly changes the energy dependence of the TDOS due to the spin polarization of the
tunneling electrons. This effect is described in Sec.~\ref{sec:TDOS}.

\section{Model}
\label{sec:model}

Technically, we employ the replicated zero-dimensional $\sigma$ model\cite{Finkelstein} in imaginary time $\tau$. This
model is formulated for calculating the disorder average of the $n$th power ($n$ is the number of replicas) of the
partition function, $\left<\mathcal{Z}^n\right>$. The standard representation of the partition function\cite{NO} is
given in terms of the coherent-state functional integral, $\mathcal{Z}^n = \int D\Psi^* D\Psi\,
e^{-\mathcal{S}[\Psi^*,\Psi]}$, with the action
\begin{widetext}
\begin{equation}
 \mathcal{S}[\Psi^*,\Psi]
  = \sum_{a=1}^n \int_0^{1/T} d\tau \left\{
      \int d\mathbf{r}\, \Psi^{+} \left[
        \frac{\partial}{\partial\tau} + H + \hat\tau_3 \Bigl(
      \xi + U_\mathrm{imp}(\mathbf{r})
    \Bigr)
      \right] \Psi
      +\frac{e^2}{2C} \left[
        \int d\mathbf{r}\, \Psi^{+} \hat\tau_3 \Psi
      \right]^2
    \right\}.
 \label{action_psi}
\end{equation}
\end{widetext}
The fermionic two-component field $\Psi = \{\psi_\uparrow, \psi^*_\downarrow\}$ consists of Grassmann (anticommuting)
variables dependent on the space coordinate $\mathbf{r}$, imaginary time $\tau$, and replica index $a$. The
two-component structure of $\Psi$ corresponds to the Nambu-Gor'kov representation, which we need for studying
superconductive correlations induced in the normal grain by the proximity to the superconductor. The other notations
used in Eq.\ (\ref{action_psi}) are $\xi = (-i \nabla)^2/2m - \mu$, $U_\mathrm{imp}(\mathbf{r})$ is the potential of
impurities, and $\hat\tau_i$ are the Pauli matrices in the Nambu-Gor'kov domain. The contact to the lead is described by
a tunnel term, which we will add to the action later.

The action (\ref{action_psi}) contains the fourth-order term due to the Coulomb interaction. The disorder averaging (we
assume Gaussian $\delta$-correlated disorder) induces another quartic term that mixes different replicas. These two
terms are decoupled by the Hubbard-Stratonovich transformation with the help of a scalar variable $\phi^a_\tau$ and a
matrix field $Q^{ab}_{\tau\tau'}$. These two objects are determined in the space of replicas and imaginary times (or,
equivalently, Matsubara energies). $Q$ is also a $2\times2$ matrix in the Nambu-Gor'kov space. After the
Hubbard-Stratonovich transformation, the action becomes quadratic in $\Psi$ and the Gaussian integration yields
\begin{multline}
 \mathcal{S}[Q,\phi]
  = \frac{\pi\nu}{4\tau_\mathrm{imp}} \TR Q^2
    + \sum_a \int_0^{1/T} d\tau\; \frac{(\phi^a_\tau)^2}{4E_C} \\
    - \TR \ln \left[
      \xi - i\hat\tau_3(\varepsilon + iH) - i\phi - \frac{iQ}{2\tau_\mathrm{imp}}
    \right].
 \label{action_log}
\end{multline}
Here $\varepsilon = i\partial/\partial\tau$ is the Matsubara energy and $\tau_\mathrm{imp}$ is the mean free time. The
``$\TR$'' symbol stands for the trace in all the three spaces of replicas, energies, and Nambu-Gor'kov, along with the
integration in the real space. We assume the grain to be so small that the magnetic field has only a Zeeman but not
orbital effect, $d \ll \Phi_0/\sqrt{D E_g}$. The zero-dimensional approximation ($d < \sqrt{D/\Delta}$) also implies
that both $Q$ and $\phi$ do not vary in space and thus commute with $\xi$.

In the derivation of the model (\ref{action_log}) we took advantage of the homogeneity of the magnetic field. The
direction of the field is chosen to be the spin quantization axis. In this particular representation, the interaction
with the field has only diagonal matrix elements within the Nambu-Gor'kov domain [see Eq.\ (\ref{action_psi})]. In the
situation of any other direction of spin quantization or inhomogeneous magnetic field a more general model is needed
with the twice larger $Q$ matrix bearing also the spin indices. In our situation this spin-dependent $Q$ matrix is
reduced to a block-diagonal form describing the ``up'' and ``down'' spin states separately. The action
(\ref{action_log}) determines the full dynamics of one of these blocks (``up''). The action for the second block differs
from Eq.\ (\ref{action_log}) only by the sign of $H$. Therefore, a physical quantity is given by the average of the two
values calculated with the action (\ref{action_log}) at $\pm H$. Having in mind this recipe, we derive all further
results from the simplified model with the action (\ref{action_log}).

The $\sigma$ model derivation proceeds with the expansion of the logarithm in Eq.\ (\ref{action_log}) in soft modes of
the $Q$ field.\cite{Efetov} These modes are concentrated at small energies, $|\varepsilon| < 1/\tau_\mathrm{imp}$, and
lie on the manifold $Q^2 = 1$. At higher energies the matrix is diagonal, $Q = \hat\tau_3 \sgn\varepsilon$. Before
expanding the logarithm, we have to exclude high-energy modes, associated with the fluctuations of the chemical
potential, from the $Q$ matrix. This is achieved by the gauge transformation\cite{KA}
\begin{equation} \label{gauge}
 Q^{ab}_{\tau\tau'}
  = e^{i\hat\tau_3 K^a_\tau} \tilde Q^{ab}_{\tau\tau'} e^{-i\hat\tau_3 K^b_{\tau'}},
\end{equation}
with properly chosen phase $K^a_\tau$. The matrix $Q^{ab}_{\tau\tau'}$ depends on the two imaginary time indices and is
\emph{antiperiodic} on the interval $[0,1/T]$. The gauge transformation may not alter these boundary conditions; thus,
we have to impose the restriction
\begin{equation}
 K^a_{1/T} - K^a_0
  = 2 \pi W^a,
 \label{W}
\end{equation}
with arbitrary integer $W^a$. For a more rigorous calculation one also has to take into account the half-integer values
of $W^a$. These values of the winding numbers imply that $\tilde Q$ is \emph{periodic} with respect to both imaginary
time indices. The half-integer $W^a$ are responsible for the parity effect.\cite{AN} However, this effect is extremely
weak in the proximity structure. To observe the parity effect, the two conditions should be fulfilled: (i)~the minigap
is of the order of $E_C$ and (ii)~the system is in the strong Coulomb blockade regime\cite{OSF} in which the ``Coulomb
staircase'' is well pronounced. These two requirements are strongly inconsistent due to the large value of the
junction's conductance $G$. Thus hereafter we consider only the integer $W^a$.

Assuming that $\tilde Q$ contains only soft modes we expand the logarithm and obtain\cite{KA}
\begin{multline}\label{action_phi}
 \mathcal{S}[\tilde Q,\phi]
  = -\frac{\pi}{\delta}\Tr \left[
      \bigl( (\varepsilon+iH) \hat\tau_3 + \phi - \dot K \bigr) \tilde Q
    \right] \\
    + \sum_a \int_0^{1/T} d\tau \left[
      \frac{(\phi^a_\tau)^2}{4E_C}
      + \frac{(\phi^a_\tau - \dot K^a_\tau)^2}{\delta}
    \right].
\end{multline}
Here the ``Tr'' operation implies trace in all the three domains of the $Q$ field. The last term of Eq.\
(\ref{action_phi}) comes from the logarithm expansion at energies well above $1/\tau_\mathrm{imp}$. It corresponds to
the static compressibility of the electron gas. Thus we have to choose $K$ such that $|\phi - \dot K|$ is minimized. The
electric potential $\phi^a_\tau$ is a real Bose field, $\phi^a_\tau = T\sum_\omega \phi^a_\omega e^{-i\omega\tau}$. We
separate the zeroth Fourier component into the integer and fractional parts, $\phi^a_{\omega=0} = 2\pi(W^a + w^a)$, and
choose $K^a_\tau$ to be
\begin{equation}
 K^a_\tau
  = C^a
    + 2 \pi T W^a \tau
    - T\sum_{\omega \neq 0}
      \frac{\phi^a_\omega}{i\omega}\,e^{-i\omega\tau}.
 \label{K}
\end{equation}
The gauge transformation with such a definition of the phase $K$ was proposed in Ref.~\onlinecite{ET}. Note that we
still have the freedom of adding an arbitrary time-independent constant $C^a$ to $K$. Indeed, this will neither change
$|\phi - \dot K|$ nor violate the restriction~(\ref{W}). The value of this constant will be fixed later after we add the
boundary term to the action.

Now we can rewrite the action~(\ref{action_phi}) in terms of $K$ getting rid of the potential $\phi$. Then the
functional integration over $\phi$ is replaced by integration over $K$ restricted by Eq.\ (\ref{W}) along with summation
over $W$ and integration over $w$ in the interval $[-1/2,1/2]$. We also use the inequality $E_C \gg \delta$ and obtain
\begin{multline}
 \mathcal{S}[\tilde Q, K]
  = -\frac{\pi}{\delta} \Tr \left[
      \bigl(
        (\varepsilon + iH) \hat\tau_3 + 2\pi T w
      \bigr) \tilde Q
    \right]\\
    +4\pi^2 T \sum_a \left[
      \frac{(w^a)^2}{\delta} + \frac{W^a w^a}{2E_C}
    \right]
    +\sum_a \int_0^{1/T} d\tau\,
      \frac{(\dot K^a_\tau)^2}{4E_C}.
 \label{action_w}
\end{multline}

Up to now we have not taken into account the tunneling of electrons to and from the superconducting lead attached to the
grain. The action (\ref{action_w}) has only one nontrivial, but still diagonal in Nambu-Gor'kov space, term (the one
containing $\hat\tau_3$). The coupling to the superconductor will induce off-diagonal contributions as well. To derive
the $\sigma$ model with the boundary term\cite{Efetov} one has to add the tunneling term to the single-particle
Hamiltonian. This term will appear in the argument of the logarithm in Eq.\ (\ref{action_log}). Expansion to the second
order in the tunneling amplitude then leads to an additional contribution to the action,
\begin{equation}
 \mathcal{S}_b
  = -\frac{\pi G}{4} \Tr \left(
      Q_1 Q_2
    \right)
  = -\frac{\pi G}{4} \Tr \left(
      Q_S e^{i\hat\tau_3 K} \tilde Q e^{-i\hat\tau_3 K}
    \right),
\end{equation}
which is to be added to Eq.\ (\ref{action_w}). Generally, the boundary term contains the trace of the product of $Q$
matrices on both sides of the contact (we denote them $Q_1$ and $Q_2$). Varying the full action, including the boundary
term, with respect to $Q$, one can easily obtain the well-known Kupriyanov-Lukichev boundary conditions.\cite{KL} In the
particular case of the normal grain coupled to the superconductor, we express $Q$ in terms of $\tilde Q$ and $K$
according to Eq.\ (\ref{gauge}). In the bulk superconductive lead the $Q$ matrix takes the value
\begin{equation} \label{QS}
 Q_S
  = 2\pi \delta^{ab} \delta(\varepsilon-\varepsilon') \hat\tau_1.
\end{equation}
This form of $Q_S$ is valid at low energies $\varepsilon \ll \Delta$. Below we consider various properties of the normal
grain at energies not larger than $E_g$; thus, the above approximation is suitable for our purposes. The high-energy
contribution to the action~(\ref{action_w}), taking into account the energy dependence of $Q_S$, leads to the
renormalization of the capacitance $C = C_0 + e^2 G/2\Delta$ as described in Ref.~\onlinecite{LO}.

Throughout the paper we assume that the temperature lies in the range
\begin{equation}
 \delta \ll T \ll \tilde E_g.
 \label{T}
\end{equation}
According to Eq.\ (\ref{action_w}), the condition $T \gg \delta$ fixes $w^a = 0$. The fractional part of the winding
number freezes at the same temperature as in Ref.~\onlinecite{ET}. However, contrary to the normal granular system with
large intergrain conductance, the integer part of the winding numbers may still strongly fluctuate. The reason for this
is the unconventional Coulomb blockade effect due to the superconductivity\cite{MG} as was discussed in the
Introduction. At temperatures below $\delta$, the fractional parts of the winding numbers are no longer frozen and the
separation into $W^a$ and $w^a$ makes no sense. One should use the approach developed in
Ref.~\onlinecite{AndreevBeloborodov} instead. We do not consider this limit in the present paper.

The parameter $\tilde E_g$, appearing in Eq.\ (\ref{T}), gives the characteristic energy scale of the matrix $\tilde Q$.
The definition of $\tilde E_g$ will be given below [see Eq.\ (\ref{sys})]. In the absence of a magnetic field this
parameter was found in Ref.\ \onlinecite{OSF}. The upper bound on the temperature allows one to simplify further
formulas, replacing all sums over the Matsubara energies by the corresponding integrals. Then the $\sigma$ model action
takes the form
\begin{multline}
\label{action_tilde}
 \mathcal{S}
  = -\frac{\pi}{\delta}\Tr \left[ (\varepsilon+iH) \hat\tau_3 \tilde Q \right]
    + \sum_a \int d\tau \Biggl\{
      \frac{\bigl(\dot K^a_\tau\bigr)^2}{4E_C} \\
      - \frac{\pi E_g}\delta \tr \left [
        \tilde Q^{aa}_{\tau\tau} \bigl(
          \hat\tau_1 \cos 2K^a_\tau + \hat\tau_2 \sin 2K^a_\tau
        \bigr)
      \right]
    \Biggr\},
\end{multline}
where ``$\tr$'' denotes the trace in the Nambu-Gor'kov space. In the absence of a magnetic field and Coulomb
interaction, the action~(\ref{action_tilde}) is the same as for the superconductive grain with the order parameter $E_g
= G\delta/4$; therefore, $E_g$ plays the role of the bare minigap in our problem.\cite{GK}

Due to the condition $E_C \gg \delta$, we can employ the adiabatic approximation.\cite{OSF} Considering $K$ as a
relatively ``fast'' variable in comparison with $\tilde Q$, we integrate the action with respect to $K$ at fixed $\tilde
Q$. Then we come to the action for $\tilde Q$ only and employ the saddle-point approximation. The simplest saddle point
of that action is diagonal in replicas and Matsubara energies but not in the Nambu-Gor'kov space. Then the condition
$\tilde Q^2 = 1$ may be explicitly resolved by the parametrization
\begin{equation} \label{ansatz}
 \tilde Q^{ab}_{\varepsilon \varepsilon'}
  = 2\pi \delta^{ab} \delta(\varepsilon-\varepsilon') \left[
      \hat\tau_3 \cos\theta^a_\varepsilon + \hat\tau_1 \sin\theta^a_\varepsilon
    \right].
\end{equation}
The angle $\theta_\varepsilon$ is the standard Usadel angle.\cite{Usadel,S&M} Generally, the $\tilde Q$ matrix may also
contain a $\hat\tau_2$ component that is not present in Eq.\ (\ref{ansatz}). This term can always be eliminated by the
proper choice of the constant $C^a$ in the definition~(\ref{K}) of the phase $K$.

All eigenvalues of $\tilde Q$ are $\pm 1$. In each replica and at every Matsubara energy we have a pair of $+1$ and $-1$
eigenvalues. Generally, the condition $\tilde Q^2 = 1$ admits an arbitrary distribution of the eigenvalue signs at small
energies, $|\varepsilon| < 1/\tau_\mathrm{imp}$. However, the inequality $T \gg \delta$ allows one to neglect these
unconventional saddle points. This is provided by the very first term of the action~(\ref{action_tilde}), namely,
$-\pi\delta^{-1}\Tr(\varepsilon \hat\tau_3 \tilde Q)$. The minimal Matsubara energy is $\pi T$; therefore, the action of
those saddle points is larger at least by $2\pi^2 T/\delta$. The proximity effect, which is accounted for by other terms
of the action, makes this estimate even stronger at low energies. We will discuss this issue below.

Substituting the ansatz~(\ref{ansatz}) into Eq.\ (\ref{action_tilde}), we find out that the action for $K^a$ is local in
imaginary time. This allows us to describe the dynamics of $K^a$ by the following Hamiltonian:
\begin{equation} \label{ham}
 \hat H^a
  = E_C \left(
      - \frac{\partial^2}{\partial K^2} - 2q^a \cos 2K
    \right),
\end{equation}
where the parameter $q^a$ is determined by
\begin{equation} \label{q}
 q^a
  = \frac{\pi E_g}{2 E_C \delta}\tr \bigl( \hat\tau_1\tilde Q^{aa}_{\tau\tau} \bigr)
  = \frac{E_g}{2E_C \delta} \int_{-\Delta}^\Delta d\varepsilon \sin\theta^a_\varepsilon.
\end{equation}
We cut off the logarithmically divergent integration at $\Delta$ since expression (\ref{QS}), which we used for $Q_S$,
is valid only at $\varepsilon \ll \Delta$. The dynamics of $K$ is restricted by the condition~(\ref{W}). The summing
over all integer $W^a$ results in the periodic boundary conditions for the eigenfunctions of the Hamiltonian~(\ref{ham})
on the interval $[0,2\pi]$. This Hamiltonian has an important symmetry: it commutes with the transformation $K \mapsto K
+ \pi$. This is related to the conservation of the electron parity in the grain. The electron number operator is $\hat n
= -i\partial/\partial K$. In this representation the first term of the Hamiltonian~(\ref{ham}) is diagonal while the
second one can change $n$ by $\pm 2$ only. Physically, this property is a consequence of the Andreev reflection
mechanism that changes the charge by $\pm 2e$. Another symmetry of the Hamiltonian~(\ref{ham}) is the inversion $K
\mapsto -K$. This is due to the particle-hole symmetry. It can be lifted by an external gate, which we do not consider
in this paper.

The adiabatic approximation relies on the fact\cite{OSF} that the characteristic frequency of the phase $K$ fluctuations
is much larger than $\tilde E_g$. At the temperatures under discussion [see Eq.\ (\ref{T})], $K$ is frozen in the ground
state of the Hamiltonian (\ref{ham}) with the energy $E_0(q) = E_C a_0(q)$, where $a_0(q)$ is the zeroth Mathieu
characteristic value. The effective action for the $\tilde Q$ matrix in terms of the angle $\theta$ is then
\begin{equation} \label{action_theta}
 \mathcal{S}
  = \sum_a \int d\tau \left[
      - \frac 1{\delta} \int d\varepsilon (\varepsilon + iH) \cos \theta^a_\varepsilon
      + E_0(q^a)
    \right].
\end{equation}
The fact that the action is represented as a sum of independent identical contributions from  each replica is due to the
trivial in replicas ansatz~(\ref{ansatz}). The next step is the saddle-point approximation. The specific form of the
action~(\ref{action_theta}) implies that the saddle-point value of angle $\theta$ is independent of the replica index.
Using this fact we omit all replica indices hereafter.

The variation of Eq.\ (\ref{action_theta}) gives $\tan\theta_\varepsilon = \tilde E_g/(\varepsilon + iH)$, where the
constant $\tilde E_g$ is determined by the system of the self-consistency equations
\begin{equation} \label{sys}
 \frac{\tilde E_g}{E_g}
  = - \frac 1{2 E_C}\frac{\partial E_0}{\partial q},
 \qquad
 q
  = \frac{E_g \tilde E_g}{E_C \delta} \ln \frac{2\Delta}{\Omega(\tilde E_g,H)}.
\end{equation}
Here we introduce the notation
\begin{equation} \label{Omega}
 \Omega(\tilde E_g,H)
  = \max(\tilde E_g,H) + \sqrt{\max\nolimits^2 (\tilde E_g,H) - \tilde E_g^2}.
\end{equation}
The last equation of Eqs.\ (\ref{sys}) is obtained from Eq.\ (\ref{q}) where the found value of angle
$\theta_\varepsilon$ was substituted.

The parameter $\tilde E_g$ has the meaning of the renormalized minigap in the thermodynamic density of states for one
spin subband. The thermodynamic DOS itself (for spin up) is obtained from $\tilde Q$ after the analytic continuation to
the real energy $E$ in the following way:
\begin{equation} \label{dos_therm}
 \rho_\uparrow (E)
  = \frac 1\delta \Re \left. \tr \hat\tau_3 \tilde Q_{\varepsilon\varepsilon}
  \right|_{\varepsilon \to -iE+0} .
\end{equation}
As a result, it acquires the standard BCS form shifted by $H$ (while the DOS for the spin down, $\rho_\downarrow$, is
shifted by $-H$). The total DOS is
\begin{gather}
 \rho(E)
  = \frac{1}{2}\Bigl[
      \rho^{\mathrm{BCS}}(E + H) + \rho^{\mathrm{BCS}}(E - H)
    \Bigr], \label{thermDOS}\\
 \rho^{\mathrm{BCS}}(E)
  = \frac{2}{\delta}\Re\cos\theta_\varepsilon \Bigr|_{\varepsilon \to -iE + 0}
  = \frac{2}{\delta} \Re \frac{|E|}{\sqrt{E^2 - \tilde E_g^2}}. \label{BCSDOS}
\end{gather}

Let us now return to the discussion of unconventional saddle points. Suppose at some energy $\varepsilon$ and in a
particular replica the $\tilde Q$ matrix has equal eigenvalues and hence is proportional to the identity matrix $\hat
\tau_0$ instead of Eq.\ (\ref{ansatz}). This results in the effective exclusion of this replica-energy pair from the
action (\ref{action_tilde}), where only the combinations $\tr(\hat\tau_{1,2,3} \tilde Q)$ are present. The parameter
$q$, given by Eq.\ (\ref{q}), is also reduced by the contribution from the energy $\varepsilon$. Due to Eq.\
(\ref{action_theta}), the total change of the saddle-point action is
\begin{multline}
 \Delta \mathcal{S}
  = \frac{2\pi}{\delta}\, \varepsilon \cos\theta_\varepsilon
    -\frac{\partial E_0}{\partial q}\, \frac{\pi E_g}{E_C \delta} \sin \theta_\varepsilon \\
  = \frac{2\pi}{\delta} \sqrt{(\varepsilon + iH)^2 + \tilde E_g^2}. \label{eq22}
\end{multline}
The last identity is based on the self-consistency equation~(\ref{sys}). The minigap $\tilde E_g$ is also changed;
however, this leads to a higher-order correction in comparison with Eq.\ (\ref{eq22}). It is easy to see that the real
part of $\Delta\mathcal{S}$ is not less than $2\pi|\varepsilon|/\delta \geqslant 2\pi^2 T/\delta$ for any values of
$\tilde E_g$ and $H$. Thus the estimate based on the first term of Eq.\ (\ref{action_tilde}) becomes even stronger when
the other terms are taken into account. For example, in the case of zero magnetic field the lower bound is increased to
$2\pi \tilde E_g/\delta$. This means that the proximity effect gives an additional ground for neglecting the saddle
points of the form other than Eq.\ (\ref{ansatz}).

So far, we have found the main saddle point in the replica-trivial sector of the $\sigma$ model. This result is
equivalent to the direct calculation of the free energy of the system, averaged over disorder. Indeed, the form of the
action~(\ref{action_theta}) implies that the average partition function obeys the identity $\langle\mathcal{Z}^n\rangle
= \langle\mathcal{Z}\rangle^n$. Using this identity and putting the number of replicas to $1$, we have for the free
energy $\mathcal{F} = -T\langle\ln\mathcal{Z}\rangle = -T\ln\langle\mathcal{Z}\rangle =
T\left.\mathcal{S}\right|_{n=1}$. Finally, substituting the saddle-point solution $\theta_\varepsilon$ into the action
(\ref{action_theta}) and noting that the imaginary-time integration simply yields a $1/T$ multiplier in this expression,
we find
\begin{equation}
 \mathcal{F}
  = - \frac 1{\delta} \int \frac{d\varepsilon (\varepsilon + iH)^2}{\sqrt{(\varepsilon + iH)^2 + \tilde E_g^2}}
    + E_0(q).
\end{equation}
This free energy has the meaning of the Landau-Ginzburg functional while $\tilde E_g$ plays the role of the order
parameter. The integral in the above expression contains a divergent contribution from high energies. As in the standard
theory of superconductivity, we get rid of this divergence, subtracting the value of the free energy in the ``normal''
state, i.e., at $\tilde E_g = 0$. Then the result of integration is
\begin{multline} \label{freeen}
 \mathcal{F} - \mathcal{F}_N
  = \frac{\tilde E_g^2}\delta \left[
      \ln\frac{2\Delta}{\Omega(\tilde E_g, H)} - \frac 12
    \right] \\
    + \frac {H}{\delta} \left( H - \sqrt{ \max\nolimits^2(\tilde E_g, H) - \tilde E_g^2} \right)
    + E_0(q).
\end{multline}

The self-consistency equations~(\ref{sys}) can be obtained by varying this free energy functional. The solution of Eqs.\
(\ref{sys}) gives extrema of the free energy; in particular, the trivial solution $\tilde E_g = 0$ always satisfies
Eqs.\ (\ref{sys}). We can also estimate fluctuations near the found extremal points. The complete calculation taking
into account all possible fluctuating modes including those that break the replica symmetry is cumbersome, but leads to
a simple result: the saddle-point approximation is valid provided $\tilde E_g \gg \delta$. The details of this
calculation for zero magnetic field can be found in Ref.~\onlinecite{OF}, where it was shown that the fluctuations
produce a negligible correction to the Josephson current in the Coulomb-blockaded junction between two superconductors
via the normal-metallic grain if $\tilde E_g \gg \delta$.

\section{Thermodynamic minigap}
\label{sec:minigap}

Generally, the system of the self-consistency equations~(\ref{sys}) is not analytically solvable. Nevertheless, the
ground-state energy for the Hamiltonian~(\ref{ham}) can be explicitly found in the two limiting cases of small and large
$q$ (physically, these limits correspond to the strong and weak Coulomb blockade, respectively). Then Eqs.\ (\ref{sys})
allow an explicit solution.

\subsection{Strong Coulomb blockade} \label{sec:minigap:strong}

We start with the case of strong Coulomb blockade. This limit implies $q \ll 1$, which means that the Coulomb energy
$E_C$ is much larger than the effective Josephson energy $E_J$. The phase $K$, which is governed by the
Hamiltonian~(\ref{ham}), is delocalized and strongly fluctuates. At the same time, the charge of the grain is almost
fixed. The Coulomb blockade wins the competition versus the proximity effect; the minigap is exponentially suppressed.

At $q\ll 1$, the potential energy in the Hamiltonian~(\ref{ham}) can be considered as a perturbation. The perturbation
theory yields the ground-state energy $E_0(q) = -E_C q^2/2$. Then the self-consistency equations~(\ref{sys}) lead to
\begin{equation} \label{sc}
 \frac{2 E_C \delta}{E_g^2}
  = \ln \frac{2\Delta}{\Omega(\tilde E_g,H)}.
\end{equation}
Nonzero solutions exist only below the following value of the magnetic field:
\begin{equation} \label{HcS}
H_c^\mathrm{S} = 2\Delta \exp\left( - \frac{2 E_C \delta}{E_g^2} \right).
\end{equation}
The quantity $\Omega(\tilde E_g,H)$ is defined by Eq.\ (\ref{Omega}) and has different meanings for $H$ below and above
$\tilde E_g$. As a result, at $H<H_c^\mathrm{S}$, Eq.\ (\ref{sc}) has two nonzero solutions corresponding to these two
cases:
\begin{equation} \label{Eg}
 \tilde E_g = \begin{cases}
 H_c^\mathrm{S} &\text{at } H< H_c^\mathrm{S}, \\
 \sqrt{H_c^\mathrm{S} (2H -H_c^\mathrm{S})} &\text{at } H_c^\mathrm{S}/2 < H < H_c^\mathrm{S}.
 \end{cases}
\end{equation}
Here the conditions for the two branches, $H< \tilde E_g(H)$ and $H> \tilde E_g(H)$, are rewritten in terms of the fixed
value $H_c^\mathrm{S}$. The double-valued structure of the whole solution becomes clear in this representation.

\begin{figure}
 \includegraphics[width=\columnwidth]{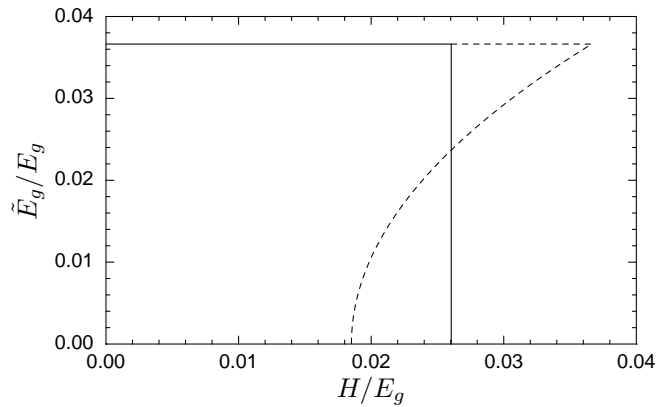}
\caption{$\tilde E_g(H)$ dependence in the limit of the strong Coulomb blockade. The solid line corresponds to the value
of $\tilde E_g$ that gives the absolute minimum to the free energy. The first-order phase transition occurs at $H =
H_c^{\text{I}}$, where $\tilde E_g$ vanishes abruptly. The dashed line shows other solutions of the self-consistency
equations~(\protect\ref{sys}). These extra solutions exist only in the interval $H_c^{\text{N}} < H < H_c^{\text{S}}$,
where the gapped and gapless states coexist. The curves are plotted for $E_C\delta/E_g^2 = 4.5$. Other parameters are $G
= 40$ and $\Delta/E_g = 150$.}
 \label{fig:gap_sC}
\end{figure}

One can easily see that the first branch in Eq.\ (\ref{Eg}), which we call the gapped (S) state, corresponds to a local
minimum of the free energy (\ref{freeen}), while the second branch gives a maximum. The gapless (N) state $\tilde E_g =
0$ [this trivial solution of Eq.\ (\ref{sc}) exists at any $H$] minimizes the free energy if $H
> H_c^\mathrm{S}/2$ and maximizes it otherwise. The fields $H_c^\mathrm{S}$ and $H_c^{\text{N}} = H_c^\mathrm{S}/2$ are
the absolute instability fields for the S and N states, respectively. In the interval $H_c^{\text{N}} < H <
H_c^{\text{S}}$, the two minima of the free energy coexist. At some value of magnetic field $H_c^{\text{I}}$ lying in
this interval the energies of the two states are equal. This is the phase equilibrium point where the first-order phase
transition occurs. Using the free energy (\ref{freeen}), we find this critical field:
\begin{equation}
 H_c^{\text{I}}
  = \frac{H_c^\mathrm{S}}{\sqrt{2}}.
\end{equation}
The $\tilde E_g(H)$ dependence is illustrated in Fig.~\ref{fig:gap_sC}.

The mechanism underlying the first-order phase transition between the gapped and gapless state is exactly the same as in
a bulk ferromagnetic superconductor.\cite{Sarma} The exchange field of the ferromagnet plays the same role as the
magnetic field in our case. The correspondence becomes complete if the superconductive pairing constant is taken to be
$\lambda = E_g^2 / 2E_C \delta$, the Debye cutoff $\omega_D$ is replaced by $\Delta$ [see Eq.\ (\ref{HcS})], and the
order parameter is $\tilde E_g$. Then the critical magnetic field $H_c^{\text{I}}$, at which the first-order phase
transition occurs, is simply the Clogston--Chandrasekhar critical field.\cite{CC}

\subsection{Weak Coulomb blockade}

Now we turn to the opposite limit of a weak Coulomb blockade and large $q$, which means that the charging energy $E_C$
is much smaller than the effective Josephson energy $E_J$. The cosine potential in the Hamiltonian (\ref{ham}) strongly
localizes the phase $K$ near $0$ and $\pi$ values. At the same time, the fluctuations of charge are strong. The
proximity effect wins against the Coulomb blockade and the minigap is only slightly suppressed in comparison with its
bare value $E_g$.

\begin{figure}
 \includegraphics[width=\columnwidth]{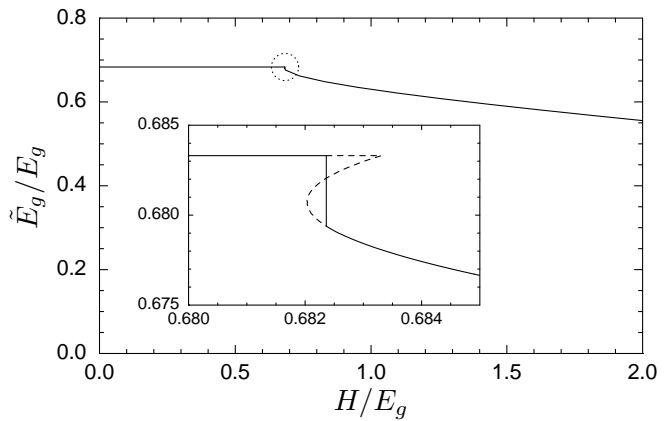}
\caption{$\tilde E_g(H)$ dependence in the limit of the weak Coulomb blockade. The inset is a close-up of the $H =
\tilde E_g(0)$ point that is encircled in the main plot. As in Fig.~\protect\ref{fig:gap_sC}, the solid line shows the
value of $\tilde E_g$ that gives the absolute minimum to the free energy, while the dashed line corresponds to extra
solutions of the self-consistency equations~(\protect\ref{sys}). The curves are plotted for $E_C\delta/E_g^2 = 1.5$.
Other parameters are $G = 40$ and $\Delta/E_g = 150$.}
 \label{fig:gap_wC}
\end{figure}

To solve Eqs.\ (\ref{sys}), we approximate the deep minima of the $\cos 2K$ potential by a one-dimensional oscillator
with the ground-state energy $E_0 (q) = -2E_C (q - \sqrt{q})$. Then, solving Eqs.\ (\ref{sys}), we find a small
correction to the bare minigap:
\begin{equation} \label{tEg2}
 \tilde E_g
  = E_g - \frac 12 \sqrt{\frac{E_C \delta}{\ln \bigl( 2\Delta / \Omega(E_g,H) \bigr)}}.
\end{equation}
This dependence is again qualitatively different for magnetic fields above and below $\tilde E_g$ (approximately equal
to $E_g$). At small magnetic field the minigap does not depend on $H$. The state with the field-independent minigap,
coinciding with the zero-field value $\tilde E_g(0)$, is similar to the S state in the strong-Coulomb-blockade regime.
At higher fields $\tilde E_g$ is logarithmically diminished. This state will be referred to as S$'$. A more accurate
analysis is needed to investigate the vicinity of the $H = \tilde E_g$ point. It turns out that the first-order phase
transition found in the opposite limit of strong Coulomb blockade persists. However, now the minigap $\tilde E_g$, being
independent of magnetic field at low $H$, experiences a very small steplike decrease and then gradually diminishes. This
is the first-order transition S--S$'$. The $\tilde E_g(H)$ dependence is shown in Fig.~\ref{fig:gap_wC}, where the inset
illustrates details near the $H = \tilde E_g(0)$ point.

From the free energy (\ref{freeen}), we straightforwardly calculate all details of the first-order phase transition,
which now occurs between two different gapped states, in the limit $q \gg 1$. We omit this bulky calculation and give
the results only. The magnetic field $H_c^{\text{S}}$, at which the S state becomes absolutely unstable, is
$H_c^{\text{S}} = \tilde E_g(0)$. The field of absolute instability of the S$'$ state we denote by $H_c^{\text{S}'}$.
Along with the critical field $H_c^{\text{I}}$ they are
\begin{gather}
 H_c^{\text{S}'}
  = \tilde E_g(0) - \frac{E_g}{2} x,
\qquad
 H_c^{\text{I}}
  = \tilde E_g(0) - \frac{E_g}{3} x, \\
 x
  = \frac{E_C\delta}{16 E_g^2 \ln^3(2\Delta/E_g)},
\end{gather}
where $\tilde E_g(0)$ is taken from Eq.\ (\ref{tEg2}). The parameter $x$, which determines the scale of the phase
coexistence region, is linear in small $E_C$ but contains also an enormously small numerical coefficient. This is the
reason why this region is extremely small in Fig.~\ref{fig:gap_wC}.

When the first-order transition occurs, the free energy has two minima with identical values. What is the energy barrier
between these two minima? This barrier is also numerically very small and equals $2 E_g^2 x^{3/2} / 3\delta$. Obviously,
when the height of this barrier becomes comparable with the temperature, fluctuations smear the first-order transition.
Then a crossover between S and S$'$ states occurs instead of a phase transition.

Finally, when the magnetic field is high enough (beyond the scope of our model), it suppresses the superconductivity in
the lead. In the absence of Coulomb effects, the minigap persists as long as the lead is superconducting and disappears
at the critical field $H_{c2}$ of the lead. If the weak Coulomb blockade is realized at $H=0$, then the minigap will
vanish at a field slightly smaller than $H_{c2}$.

\subsection{Intermediate case}

\begin{figure}
 \includegraphics[width=\columnwidth]{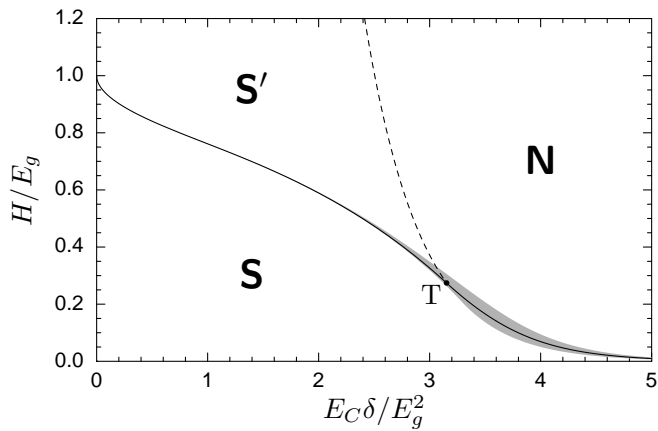}
\caption{Phase diagram $E_C$--$H$. The solid line marks the first-order transition from the S to either N or S$'$ state.
The region of phase coexistence is shaded. The dashed line shows the second-order S$'$--N transition. All the three
phases equilibrate at the triple point T. A detailed diagram in the vicinity of the triple point is shown in
Fig.~\protect\ref{fig:diagram_cup}. The diagram is plotted for $G = 40$ and $\Delta/E_g = 150$.}
 \label{fig:diagram}
\end{figure}

In this section we consider an intermediate regime when the Coulomb interaction is comparable with the proximity
coupling. In Fig.~\ref{fig:diagram} we present the phase diagram in the $E_C$--$H$ plane. This diagram covers all
limiting regimes considered above along with the intermediate region.

We have already studied the first-order transition from S to N and S$'$ states at large and small $E_C$, respectively.
At the same time, the line of absolute instability of the S phase can be extracted from results of Ref.~\onlinecite{OSF}
obtained at $H=0$. Indeed, at low magnetic field $H < \tilde E_g$, the self-consistency equations~(\ref{sys}) do not
contain $H$. Hence the minigap is independent of the magnetic field (S state) and coincides with the zero-field value
$\tilde E_g(0)$. The maximal possible magnetic field for this state is $H_c^{\text{S}} = \tilde E_g(0)$. Obviously, this
result holds for any value of $q$. The $\tilde E_g(E_C)$ dependence in the absence of magnetic field was studied in
Ref.~\onlinecite{OSF}.

Now we concentrate on the S$'$--N second-order transition and the vicinity of the triple point where all three phases
equilibrate. The S$'$--N transition line can be calculated analytically. Any solution of the self-consistency
equations~(\ref{sys}) gives an extremum of the free energy: $\partial \mathcal{F} / \partial \tilde E_g = 0$. The normal
state ($\tilde E_g =0$) always satisfies this condition. The normal state is stable provided $\partial^2 \mathcal{F} /
\partial \tilde E_g^2 > 0$. Calculating the second derivative of the free energy (\ref{freeen}) and then taking the limit
$\tilde E_g \to 0$, we easily find the critical magnetic field
\begin{equation} \label{HcN}
 H_c^{\text{N}}
  = \Delta \exp\left( - 2 E_C \delta / E_g^2 \right).
\end{equation}
This critical field determines the boundary of the normal region in the phase diagram in Fig.~\ref{fig:diagram}. What
happens just below this boundary? The second derivative of the free energy becomes negative. If, at the same time, the
fourth derivative is positive, then the free energy achieves a minimum at small $\tilde E_g$. This is the second-order
phase transition from the N to S$'$ state. Otherwise, if the fourth derivative is also negative, then below
$H_c^{\text{N}}$ a minimum at small $\tilde E_g$ vanishes and the only stable state has finite value of the minigap
$\tilde E_g(0)$. Thus $H_c^{\text{N}}$ is the normal-state absolute instability field for the N--S first-order
transition. In the strong-Coulomb-interaction limit, the critical field (\ref{HcN}) coincides with $H_c^{\text{N}} =
H_c^{\text{S}}/2$, which we found in Sec.~\ref{sec:minigap:strong}.

\begin{figure}
 \includegraphics[width=\columnwidth]{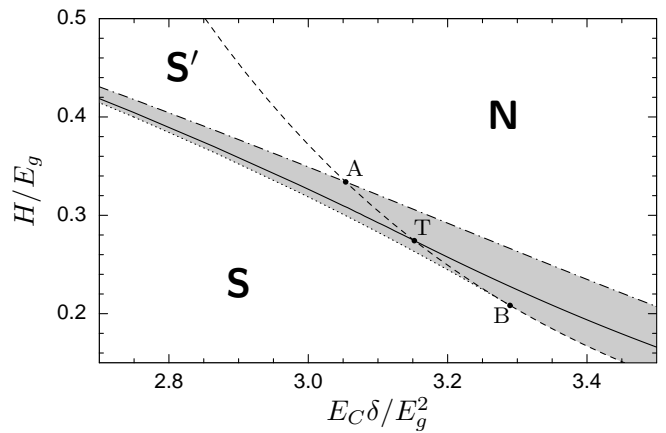}
\caption{Phase diagram near the triple point. The solid line shows the first-order transition from S to N for $E_C$
above the triple point and from S to S$'$ otherwise. The dashed line corresponds to the critical field $H_c^{\text{N}}$,
given by Eq.\ (\protect\ref{HcN}). The dotted line denotes the absolute instability of the S$'$ phase at the field
$H_c^{\text{S}'}$. This line ends at the point marked as B. The dash-dotted line is $H_c^{\text{S}}$. It intersects with
$H_c^{\text{N}}$ at the A point. The region of a possible metastable state is shaded. The diagram is plotted for $G =
40$ and $\Delta/E_g = 150$.}
 \label{fig:diagram_cup}
\end{figure}

The point on the critical line~(\ref{HcN}), where the fourth derivative of the free energy changes its sign, is denoted
as B (see Fig.~\ref{fig:diagram_cup}). To find this point, one should use a more precise value of the ground-state
energy of the Hamiltonian (\ref{ham}), taking into account the fourth-order perturbation correction: $E_0(q) =
E_C(-q^2/2 + 7q^4/128)$. Then taking the fourth derivative of the free energy (\ref{freeen}), we find the equation
$H_B^2 = E_g^4 / 7 E_C \delta$, which, together with Eq.\ (\ref{HcN}), determines the position of the B point.

In Fig.~\ref{fig:diagram_cup} a close-up of the triple-point region is shown. The $H_c^{\text{N}}$ curve is shown by the
dashed line. The line of absolute instability of the S$'$ phase $H_c^{\text{S}'}$ (the dotted line in
Fig.~\ref{fig:diagram_cup}) ends in the B point. Indeed, the S$'$ phase with arbitrary small $\tilde E_g$ exists only if
the fourth derivative of the free energy is positive. Another feature of the phase diagram, the A point, is the point
where the N--S$'$ second-order transition and the absolute instability of the S phase occur simultaneously. Finally,
between the A and B points on the $H_c^{\text{N}}$ line the triple point T lies. This is the point where the first- and
second-order transition lines intersect. All three phases have the same energy in this point.

\begin{figure}
 \includegraphics[width=\columnwidth]{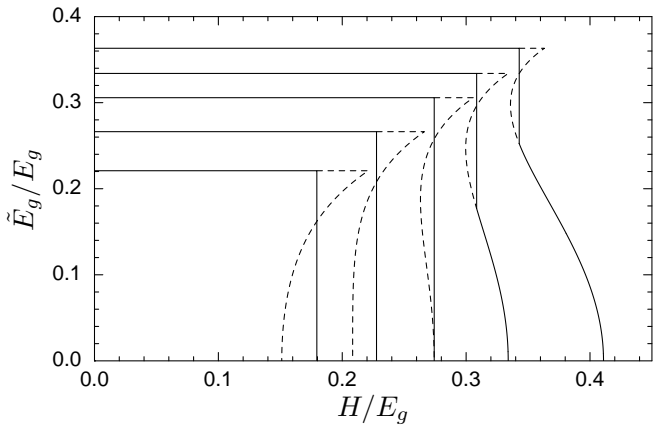}
\caption{Minigap as a function of magnetic field (solid lines) for several values of $E_C$ near the triple point. The
dashed lines show extra solutions of the self-consistency equations~(\ref{sys}). The curves from left to right
correspond to $E_C\delta/E_g^2 = 3.45$, $3.289$ (B point), $3.152$ (T point), $3.054$ (A point), and $2.95$. The other
parameters are $G = 40$ and $\Delta/E_g = 150$.}
 \label{fig:tEg}
\end{figure}

Another illustration of the complicated phases structure near the triple point is given by Fig.~\ref{fig:tEg}, where
several $\tilde E_g(H)$ dependences are shown. The leftmost curve corresponds to $E_C$ above the B point. Qualitatively,
this case is similar to the strong-Coulomb-blockade limit (see Fig.~\ref{fig:gap_sC}). The next curve is plotted for
$E_C$ taken at the B point. It looks much the same, but the unstable solution (dashed line) vanishes as the fourth
rather than square root of the magnetic field. The next curve is for the T point. The first- and second-order
transitions (solid and dashed lines) occur at the same magnetic field. The right but one curve is for the A point
[$H_c^{\text{N}} = \tilde E_g(0)$]. The rightmost curve illustrates the case of $E_C$ below the A point. It is similar
to the weak-interaction limit (see Fig.~\ref{fig:gap_wC}). The minigap vanishes continuously at $H_c^{\text{N}}$. As
$E_C$ further decreases, this critical field grows exponentially and rapidly goes beyond the scope of our model.

\section{Tunneling density of states}
\label{sec:TDOS}

Measuring the thermodynamic density of states is experimentally complicated due to small size of the sample. The
tunneling technique is more practical in this case. The actual measured quantity is the tunnel current which depends on
the voltage applied between the system and a normal-metallic external tip. The differential conductance $dI/dV$
extracted from this experiment is proportional to the local \emph{tunneling} density of states at energy $eV$. The
latter is determined by the imaginary part of the one-particle Green function. Without an interaction, the thermodynamic
and tunneling DOS coincide. However, in an interacting system the Green function is ``dressed'' by the interaction that
yields the difference between the two quantities. In $\sigma$-model language, the thermodynamic DOS is determined by the
$\tilde Q$ matrix [see Eq.\ (\ref{dos_therm})], while the tunneling DOS is given by a similar expression with the
``dressed'' matrix $Q$.

In fact, the differential conductance gives the tunneling density of states only at zero temperature. If $T > 0$, the
tunneling electrons are dispersed in the energy range of the order of $T$. The expression for $dI/dV$ then reads
\begin{equation}
 \frac{dI}{dV}
  = \frac{\delta}{2R_T} \int dE \frac{\rho^{\mathrm{tun}}(E + eV)}{4T \cosh^2(E/2T)},
\end{equation}
with $R_T$ being the tunnel resistance between the tip and the grain. In order to measure the subtle structure of the
tunneling DOS due to the Zeeman splitting, the temperature should be low enough:
\begin{equation}
 T \ll \min( \tilde E_g, H ).
\end{equation}

To calculate the TDOS for one spin projection (spin up), we should analytically continue the expression
\begin{multline}
 \rho_\uparrow^\mathrm{tun} (\varepsilon)
  = \frac 1\delta \tr \left< \hat\tau_3 Q_{\varepsilon\varepsilon} \right> \\
  = \frac 1\delta \int_{-\infty}^\infty d\tau e^{i\varepsilon\tau}
    \tr \left<
      \hat\tau_3 e^{i\hat\tau_3 K_\tau} \tilde Q_{\tau 0} e^{-i\hat\tau_3 K_0}
    \right>
\end{multline}
to the real energies, $i\varepsilon \to E + i0$, and take its real part. The angular brackets denote averaging with
weight $e^{-\mathcal S[\tilde Q, K]}$ where the action is given by Eq.\ (\ref{action_tilde}). The TDOS for the spin down
is then obtained after inverting the sign of the magnetic field. The adiabatic and saddle-point approximations allow one
to substitute $\tilde Q$ of the form~(\ref{ansatz}) and average over $K$. This procedure leads to the expression
\begin{equation}
 \rho_\uparrow^\mathrm{tun} (\varepsilon)
  = \frac 2\delta \int_{-\infty}^\infty \frac{d\omega}{2\pi}
    \frac{(\omega+iH)}{\sqrt{(\omega+iH)^2+\tilde E_g^2}} C(\varepsilon-\omega),
 \label{convolute}
\end{equation}
where $C(\omega)$ is the phase correlator containing average over the ground state of the Hamiltonian (\ref{ham}):
\begin{equation}
 C(\omega)
  = \int_{-\infty}^\infty d\tau e^{i\omega\tau} \left< \cos(K_\tau -K_0 ) \right>.
\end{equation}
This quantity can be expressed in terms of the eigenvalues $E_n$ and the eigenfunctions $\left| n \right>$ of the
Hamiltonian~(\ref{ham}):
\begin{gather}
 C(\omega)
  = \sum_n P_n \frac{2 A_n}{\omega^2 + A_n^2},
\qquad
 A_n
  = E_n - E_0, \label{C} \\
 P_n
  = \left| \left< 0 \left| \cos K \right| n \right> \right|^2
    + \left| \left< 0 \left| \sin K \right| n \right> \right|^2.
\end{gather}

\begin{figure}
\vspace{6pt}
 \includegraphics[width=\columnwidth]{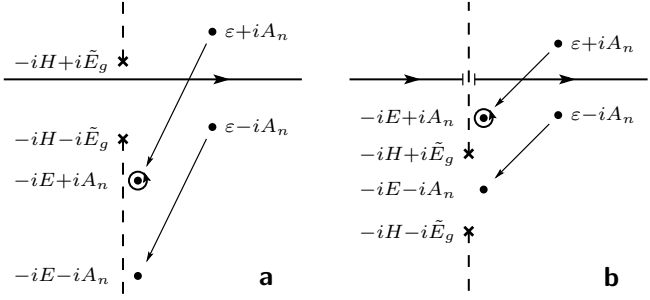}
\caption{Analytic continuation of the integrand of Eq.\ (\protect\ref{convolute}) in the complex $\omega$ plane. The
integral is taken along the real axis denoted by the line with an arrow. The dashed lines are branch-cut
discontinuities. The small dots show the poles of the $n$th term in the sum~(\protect\ref{C}) for $C(\varepsilon -
\omega)$. These poles move to new positions at the imaginary axis due to the analytic continuation. The residue in the
pole that traversed the real axis provides the real part of the integral and determines the TDOS. (a)~Left panel
corresponds to the S phase, $H < \tilde E_g$. The Coulomb and proximity gaps add up around $E = 0$. (b)~Right panel
corresponds to the S$'$ phase, $H
> \tilde E_g$, where the two gaps are separated.}
 \label{fig:analyt_continue}
\end{figure}

Now we can perform the analytic continuation to real energies in Eq.\ (\ref{convolute}). In
Fig.~\ref{fig:analyt_continue} we plot the integration contour in the complex plain of the $\omega$ variable and show
the two poles of $C(\varepsilon - \omega)$ along with the branch cut for the rest of the integrand. For simplicity we
retain only a pair of poles corresponding to the $n$th term of the sum~(\ref{C}). The analytic continuation moves the
poles; as a result, all the singularities of the integrand reside on the imaginary axis. Therefore the integral along
the real axis becomes purely imaginary and does not contribute to the TDOS. The real part --- and hence the TDOS --- is
nonzero if a pole traverses the real axis while moving (see Fig.~\ref{fig:analyt_continue}). Then the residue in this
pole will determine the result. Note that the value of this residue is nothing but the thermodynamic DOS at the energy
corresponding to the final position of the pole. To obtain the complete expression for the TDOS, we sum the
contributions from all the terms in the sum~(\ref{C}) and symmetrize the result with respect to the spin direction:
\begin{multline}
 \rho^\mathrm{tun}(E)
  = \frac 12 \Bigl[ \rho_\uparrow^\mathrm{tun}(E) + \rho_\downarrow^\mathrm{tun}(E) \Bigr] \\
  = \sum_n P_n \vartheta(|E| - A_n) \rho(|E| - A_n).
 \label{DoS}
\end{multline}
Here $\vartheta(x)$ is the Heaviside step function and $\rho(E)$ is the thermodynamic DOS given by Eq.\
(\ref{thermDOS}).

\begin{figure}
\vspace{6pt}
 \includegraphics[width=\columnwidth]{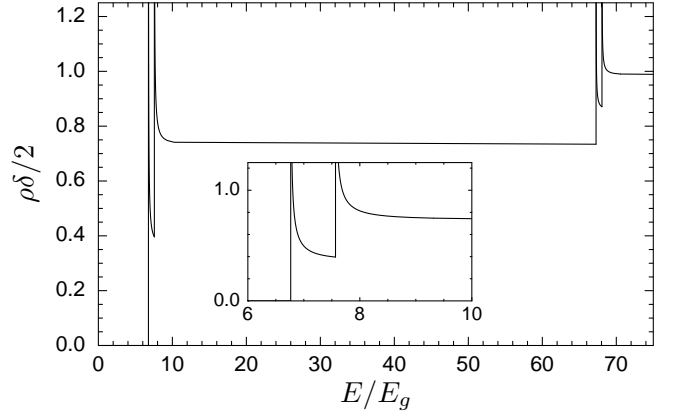}
\caption{Energy dependence of the tunneling DOS for the S phase. The parameters are $E_C\delta/E_g^2 = 2.5$, $H/\tilde
E_g = 0.4$, and $\Delta/E_g = 150$. The inset shows the details of the Zeeman splitting of the first peak. The second
peak is split in the same manner.}
 \label{fig:TDOS_hlg}
\end{figure}

\begin{figure}
\vspace{6pt}
 \includegraphics[width=\columnwidth]{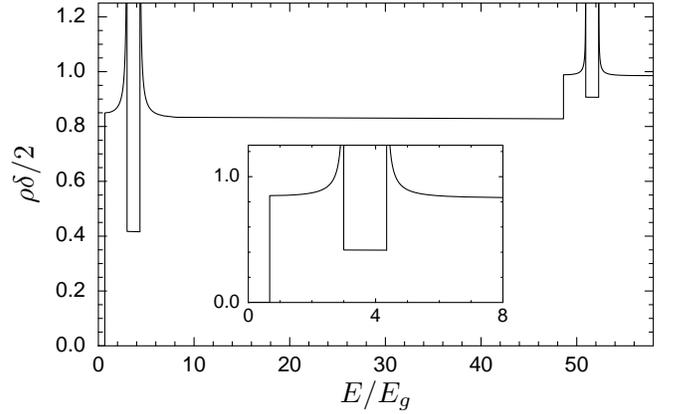}
\caption{Energy dependence of the tunneling DOS for the S$'$ phase. The parameters are $E_C\delta/E_g^2 = 1.0$,
$H/\tilde E_g = 3.0$, and $\Delta/E_g = 150$. The inset shows the details of the first peak structure. The Coulomb and
proximity gap are clearly separated for $H > \tilde E_g$.}
 \label{fig:TDOS_glh}
\end{figure}

The resulting expression for the TDOS has a clear physical meaning. Every term of the sum~(\ref{DoS}) is obtained from
$\rho(E)$ by inserting the $2A_n = 2(E_n-E_0)$ Coulomb gap around $E = 0$ and multiplying by the factor $P_n$. The
energy dependence of the TDOS is qualitatively different for the two gapped phases S and S$'$, described in the previous
section. In the S phase, when $H < \tilde E_g$, the peaks of the TDOS (see Fig.~\ref{fig:TDOS_hlg}) are split due to the
Zeeman effect, but the whole picture resembles the result of Ref.~\onlinecite{OSF}. At higher magnetic field $H > \tilde
E_g$, the grain is in the S$'$ phase. The Coulomb and proximity gaps are now separated as shown in
Fig.~\ref{fig:TDOS_glh}. The Coulomb gap is always centered around $E = 0$ while the minigap is shifted by $H$. The
density of states inside this shifted minigap is no longer zero due to the contribution from electrons with the opposite
spin.

The resulting expression~(\ref{DoS}) contains the matrix elements $P_n$ and the energy level separations $A_n = E_n -
E_0$. Both quantities can be found analytically in the limits of weak and strong Coulomb interaction. Previously, we
have mentioned two symmetries of the Hamiltonian~(\ref{ham}): it commutes with operations $K \mapsto -K$ and $K \mapsto
K + \pi$. Therefore, the $P_n$ coefficients are nonzero only for $n=4k+1$ and $n=4k+2$. Another of their properties is
the normalization $\sum_n P_n = 1$, thus $\rho(E) = \rho^\mathrm{tun}(E) = 2/\delta$ if the energy $E$ is far from the
Fermi energy. The sequence $P_n$ rapidly decreases at any strength of the Coulomb interaction. This allows us to keep
only $n=1$ and $n=2$ terms in the sum~(\ref{DoS}). In the weak-Coulomb-blockade limit, $q \gg 1$, we approximate the
$\cos 2K$ potential by two deep parabolic wells. The splitting of the two lowest levels is exponentially small. To the
main order in $1/\sqrt q$, we have
\begin{gather}
 P_1
  = 1-\frac 1{4\sqrt q},
\qquad
 P_2
  = \frac 1{4\sqrt q},
\\
 A_1
  = 0,
\qquad
 A_2
  = 4 E_C \sqrt q.
\end{gather}
In the opposite case of a strong Coulomb interaction $q \ll 1$, we employ the perturbation theory in $q$ and obtain
\begin{equation}
 P_{1,2}
  = \frac 12 \left( 1 \pm \frac q2 \right),
\qquad
 A_{1,2} = E_C ( 1 \mp q ).
\end{equation}

In the limit of a strong Coulomb interaction, at high magnetic fields $H > H_c^{\mathrm{I}}$, the minigap is absent and
the grain is in the N phase. In this case, the TDOS exhibits the pure Coulomb gap at energies below $E_C$ and is
constant above this gap: $\rho^{\mathrm{tun}}(E) = (2/\delta) \vartheta(|E| - E_C)$. In fact, even in the N regime a
very small minigap of the order of $\sqrt{\delta/H} \tilde E_g(0)$ [where $\tilde E_g(0)$, the minigap at $H=0$,
coincides with $H_c^\mathrm{S}$ given by Eq.\ (\ref{HcS})] persists in the TDOS due to the fluctuations, which we
neglected in this paper. The mechanism of this effect is similar to that for a superconductive grain in a strong
magnetic field described in Ref.~\onlinecite{Kee}.

\section{Conclusions}

In conclusion, we developed a self-consistent theory for proximity and charging effects in the presence of an external
magnetic field. The minigap induced in the grain shows a complicated dependence on magnetic field. Two distinct
minigapped states exist, and a first-order phase transition occurs between them. The transition to the gapless state is
of first order from the S state and of second order from S$'$. The tunneling DOS is also different in S and S$'$ states.
In high magnetic field $H > \tilde E_g$, the TDOS acquires two distinct gaps: a Coulomb gap at zero energy and a
proximity minigap shifted from zero by a magnetic field.

The systems discussed in this paper can be experimentally realized with the following parameters: the grain can be made
of a noble metal and have the form of a disk with thickness $d\sim 25$~nm and diameter an order of magnitude larger. At
the interface transparency of order $2\times 10^{-6}$ (insulating oxide barrier) we estimate $\delta \sim 5 \times
10^{-4}$~K, $E_g \sim 0.01$~K, $E_C \sim 0.1$~K, $E_J \sim 1$~K, and $G\sim 100$. The lead can be made of Nb; then
$\Delta \sim 12$~K and the magnetic fields should be smaller than 1~T.

\begin{acknowledgments}
We are grateful to I.~S.\ Beloborodov, A.~V.\ Lopatin, and V.~M.\ Vinokur for useful discussions. This research was
supported by RFBR Grants Nos.\ 04-02-16348 and 04-02-08159, the Russian Ministry of Education and Science under Contract
No.\ RI-112/001/417, and the program ``Quantum Macrophysics'' of the Russian Academy of Sciences. P.M.O.\ and Ya.V.F.\
also acknowledge support from the Dynasty Foundation. Ya.V.F.\ was also supported by RF Presidential Grant
No.~MK-3811.2005.2, the Russian Science Support Foundation, CRDF, and the Russian Ministry of Education and Science.
\end{acknowledgments}

\end{document}